\documentstyle[11pt,epsfig,amsbsy]{article}

\def\be{\begin{equation}}
\def\ee{\end{equation}}
\def\ba{\begin{eqnarray}}
\def\ea{\end{eqnarray}}

\def\ga{\mathrel{\raise.3ex\hbox{$>$\kern-.75em\lower1ex\hbox{$\sim$}}}}
\def\la{\mathrel{\raise.3ex\hbox{$<$\kern-.75em\lower1ex\hbox{$\sim$}}}}

\newcommand{\fr}[2]{\frac{#1}{#2}}

\newcommand{\oo}{\omega_{0}}
\newcommand{\oa}{\omega_{a}}
\newcommand{\eff}{\rm{eff}}

\newcommand{\Omo}{\Omega_{\rm{m} 0}}
\newcommand{\Oma}{\Omega_{\rm{m}}(a)}
\newcommand{\Omn}{\Omega_{\rm{m}}(n)}

\newcommand{\m}{\rm{m}}
\newcommand{\rad}{\rm{rad}}

\newcommand{\JCAP}{J.\ Cosmol.\ Astropart.\ Phys.}

\newcommand{\omde}{\omega_{\rm{de}}}
\newcommand{\omeff}{\omega_{\rm{eff}}}
\begin{document}

\baselineskip=16pt
\begin{titlepage}
\begin{center}

\vspace{0.5cm}

\large {\bf Constraints on scalar-tensor theories of gravity from observations}
\vspace*{5mm} \normalsize

{\bf Seokcheon Lee$^{\,1,2}$}

\smallskip
\medskip

$^1${\it Institute of Physics, Academia Sinica, \\
Taipei, Taiwan 11529, R.O.C.}

$^2${\it Leung Center for Cosmology and Particle Astrophysics, National Taiwan University, \\ Taipei, Taiwan 10617, R.O.C.}

\smallskip
\end{center}

\vskip0.6in

\centerline{\large\bf Abstract}
In spite of their original discrepancy, both dark energy and modified theory of gravity can be parameterized by the effective equation of state (EOS) $\omega$ for the expansion history of the Universe. A useful model independent approach to the EOS of them can be given by so-called Chevallier-Polarski-Linder (CPL) parametrization where two parameters of it ($\omega_{0}$ and $\omega_{a}$) can be constrained by the geometrical observations which suffer from degeneracies between models. The linear growth of large scale structure is usually used to remove these degeneracies. This growth can be described by the growth index parameter $\gamma$ and it can be parameterized by $\gamma_{0} + \gamma_{a} (1 - a)$ in general. We use the scalar-tensor theories of gravity (STG) and show that the discernment between models is possible only when $\gamma_a$ is not negligible. We show that the linear density perturbation of the matter component as a function of redshift severely constrains the viable subclasses of STG in terms of $\omega$ and $\gamma$. From this method, we can rule out or prove the viable STG in future observations. When we use $Z(\phi) =1$, $F$ shows the convex shape of evolution in a viable STG model. The viable STG models with $Z(\phi) = 1$ are not distinguishable from dark energy models when we strongly limit the solar system constraint.

\vspace*{2mm}

\end{titlepage}

\section{Introduction}
\setcounter{equation}{0}
The late-time acceleration of the cosmic expansion invokes the mysterious dark energy or the modification of the gravity theory beyond Einstein's general relativity. Although there exist a number of dark energy and modified gravity models, both can be described by the effective equation of state (EOS) $\omega$ when one considers the expansion history of the Universe. Chevallier-Polarski-Linder (CPL) parametrization $\omega = \omega_0 + \omega_a (1 - a)$ is one of the suitable candidates to describe the general $\omega$ \cite{0009008,0208512}.

However, models of different physical origins with the same background expansion history can not be separated with $\omega$. Thus, the growth of large scale structure is used as a complementary probe to segregate models \cite{0503644,0507184,0612452,07042421,07090307,08012431,08021068,08033292}.

The growth index parameter (GIP) $\gamma$ defined by $\fr{d \ln \delta_{\m}(a)}{d \ln a} \equiv \Oma^{\gamma}$ is often used to discriminate different models using the linear matter perturbations \cite{Peebles1,Peebles2,Lahav}. Although $\gamma$ is time dependent by its definition even for the simple dark energy models \cite{09051522,09061643,09072108}, the constant value of it can be well matched with some dark energy and modified gravity models \cite{0507263,0608681,0701317,07101092,08024122,08081316}. However, it can be generalized as a time dependent $\gamma(a)$ \cite{9804015,07101510,08032236,09052470,09053444,09084379}.

Behavior of $\gamma$ in the so-called Dvali-Gabadadze-Porrati (DGP) braneworld model \cite{0005016,0010186,0105068} has been widely investigated \cite{08033292,0701317,08024122,09051735,09053444,0401515}. $\gamma$ for a modification of gravitational action with a general function of the scalar curvature instead of the standard Einstein-Hilbert term, named f(R) gravity \cite{Buchdahl} has been also studied \cite{08032236,08093374,09035296,09082669}. In general, $\gamma$ has a scale dependence in this model. Also when one considers the scalar-tensor theories of gravity (STG) \cite{Bergmann,Nordtvedt,Wagoner}, one can obtain the specific $\gamma$ for the certain STG model \cite{07101510,0001066,08024196}.

There is a useful approximate solution for the growth factor within general relativity \cite{0507263} \be \delta_{\m}(a) = a e^{\int_{0}^{a} (da'/a') [\Omega_{\m}(a')^{\gamma} - 1]} \, , \label{Linderg} \ee where $\Oma$ is the matter density contrast, $\delta_{\m} = \delta \rho_{\m} / \rho_{\m}$ is the linear density perturbation of the matter, and $\gamma = 0.55 + 0.05 [ 1 + \omega(z=1)]$ for $\omega > -1$. However, the accuracy of this solution is misinformed. As either $\omega_{0}$ or $\omega_{a}$ increases, the accuracy of the approximate solution given in Eq. (\ref{Linderg}) is decreased. For example, the error of approximate solution is about $1.2 \%$ when one consider ($\oo, \oa$) $=$ ($-0.78, 0.32$). If $\oa$ increases to $0.4$, then the error becomes $2.0 \%$ for the same value of $\oo$.

There are both theoretical and phenomenological motivations for STG. The former is related to the existence a ubiquitous fundamental scalar coupled to gravity in theories which unify gravity with other interactions \cite{Fradkin,Callan,Lovelace,Green,Polchinski}. Also the dynamical equivalence between f(R) theories and a particular class of STG has been shown in the case of metric formalism \cite{Teyssandier,Magnano,9307034,0307338,09100434} as well as in the Palatini formalism \cite{0308111,0604028}. The later have several aspects. First, ``the lithium problem'' in the standard big bang nucleosynthesis (BBN) might be solved in STG due to the slower expansion than in general relativity before BBN, but faster during BBN \cite{0511693,0601299,08111845}. The weak lensing (WL) shear power spectrum in STG predicts the different one compared to in GR because they cause the different growth history of the matter \cite{0503644,0403654,0412120}. Integrated Sachs-Wolfe (ISW) effect probes modified gravities on cosmological scales through the matter potential relation \cite{9906066,08032238,SLAIP,09092045}. The crossing phantom $\omega < -1$ also can be naturally obtained in STG \cite{0504582,0606287,0610092}.

There have been a number of reconstructions of specific STG models which is consistent with known observational constraints \cite{0011115,0107386,0508542,0612569,07053586,08031106,10031686,10061246,10112915}. However, we need to reconstruct theory without any specific theory {\it a priori}. Thus, we use both background and growth history parameters ($\omega, \gamma$) to find the viable subclasses of STG.

We briefly review the basic background evolution equations of STG model in the next section. In Sec. 3, we also review the linear perturbation equations of the model. We derive the reconstruction equations for model functions $F(\phi)$ and $U(\phi)$ as a function of scale factor $a$ and check the viability of specific models in terms of parameters $\omega$ and $\gamma$ in Sec. 4. We conclude in Sec. 5. We also show the accuracy of the approximate solution in Eq. (\ref{Linderg}) for the general values of ($\oo, \oa$) and find the initial values of $\phi'$ and $U$ in the appendix.

\section{Scalar-Tensor Theories of Gravity}
\setcounter{equation}{0}

STG are described in the Jordan frame (JF) by the action \cite{Bergmann,Nordtvedt,Wagoner}
\be S = \fr{1}{16 \pi G_{\ast}} \int d^4 x \sqrt{-g} \Biggl[ F(\phi) R - Z(\phi) \nabla^{\mu} \phi \nabla_{\mu} \phi - 2 U(\phi) \Biggr] + S_{\m}(g_{\mu\nu}, \psi_{\m}) \, , \label{SST} \ee
where $G_{\ast}$ denotes the bare gravitational coupling constant which differs from the measured one, $F(\phi)$ and $Z(\phi)$ are dimensionless, and $U(\phi)$ is the potential of the scalar field $\phi$. $F(\phi)$ needs to be positive to ensure that the gravity is attractive. In the matter action $S_{\m}(g_{\mu\nu},\psi_{\m})$, the matter fields $\psi_{\m}$ is universally coupled to the metric $g_{\mu\nu}$ and all experimental data including Hubble parameter $H$ and redshift $z$ will thus have their usual interpretation in this JF \cite{0009034}.


The evolution equations in the flat Friedmann-Lemaitre-Robertson-Walker (FLRW) metric are given by
\ba && 3 F H^2  = 8 \pi G_{\ast} (\rho_{\m} + \rho_{\rad}) + \fr{1}{2} Z \dot{\phi}^2 - 3 H \dot{F} + U \, , \label{G00} \\ && 2 F \dot{H} = - 8 \pi G_{\ast} (\rho_{\m} + \fr{4}{3} \rho_{\rad}) - Z \dot{\phi}^2 - \ddot{F} + H \dot{F} \, , \label{Gii} \\ && Z \Bigl( \ddot{\phi} + 3 H \dot{\phi} \Bigr) = \fr{1}{2} F_{,\, \phi} R - \fr{1}{2} Z_{,\, \phi} \dot{\phi}^2 - U_{,\, \phi} \, , \label{phieq} \\ && \dot{\rho}_{i} + 3 H ( 1 + \omega_{i}) \rho_{i} = 0 \hspace{0.2in} {\rm with} \hspace{0.2in}  (\omega_{\m} =0 \, \rm{and} \, \omega_{\rad} = 1/3) \, , \label{rhoi} \ea where dots denote the differentiation with respect to (w.r.t) the (JF) cosmic time $t$, $F_{,\phi} = \fr{d F}{d \phi}$, and $R = 6(2H^2 + \dot{H})$ is the Ricci scalar. We limit our consideration after the matter dominated epoch and ignore the radiation component. The effective gravitational constant $G_{\eff}$ between two test masses measured in laboratory Cavendish-type experiments is given by \cite{0001066,Damour} \be G_{\eff} = \fr{G_{\ast}}{F} \Biggl(\fr{2 Z(\phi) F + 4F_{,\, \phi}^2}{2 Z(\phi) F + 3F_{,\,\phi}^2} \Biggr) \, . \label{Geff} \ee This is different from the Newton's gravitational constant $G_{N} = G_{\ast} / F$ as the inverse factor of the curvature scalar $R$. In Eq. (\ref{Geff}), $G_{\ast}/F$ comes from the exchange of a graviton between the two masses, whereas $(G_{\ast}/F) [F_{, \phi}^2/(2 Z F + 3 F_{, \phi}^2)]$ is due to the exchange of a scalar particle between them. When we use the familiar expression for Brans-Dicke (BD) representation, $F = \phi$ and $Z = \omega_{BD}/ \phi$, Eq. (\ref{Geff}) becomes $G_{\eff}^{BD} = G_{\ast} \phi^{-1} (2 \omega_{BD} + 4)/(2 \omega_{BD} + 3) = G_{\ast} \phi^{-1} (1 + \fr{4}{3} \beta^2) $ by using the conversion relation $\beta = \sqrt{\fr{3}{4 (2 \omega_{BD} + 3)}}$ between STG coupling constant and BD parameter \cite{07042421}. We can recover $G_{\eff} = G_{\ast}$ in $\beta \rightarrow 0$ ({\it i.e.} $\omega_{BD} \rightarrow \infty$) limit. $G_{\eff}$ is experimentally bounded
$G_{\eff}(z=0) - G_{N}(z=0) = 0.02 \%$ and  $\Bigl| \dot{G}_{\eff} / G_{eff} \Bigr| < 6 \times 10^{-12} yr^{-1} $ \cite{solar1,solar2,Cassini}. One can always reduce $Z(\phi)$ and $F(\phi)$ to one unknown function by a redefinition of the scalar field and thus we will consider $Z(\phi) = 1$ case. However, this parametrization $Z(\phi) = 1$ can sometimes be singular \cite{0009034}.

There are several features we need to emphasize. First, the positive energy does not imply that $\dot{\phi}^2 > 0$ due to the mixture of tensor and scalar degrees of freedom in the JF. As the same reason, the second derivative of $U$ does not give the precise value of its squared mass. Secondly, the evolution of the scalar field is determined by the effective potential $U_{\eff}(\phi) = U(\phi) - \fr{1}{2} F(\phi) R$ as shown in Eq. (\ref{phieq}). If we consider a light scalar field weakly coupled to matter ($F \sim {\cal O}(1)$ and $F_{,\, \phi} \sim 0$), then at early epoch $U_{\rm{eff}}$ is dominated by $F(\phi) R$-term. Thus, the scalar field is dynamically driven to General Relativity corresponding value ($\phi = 0$) and deviate from it at late epoch when $U_{\eff}$ is approximated by $U(\phi)$.

In order to study the cosmological dynamics of the system, it is convenient to rewrite the evolution Eqs. (\ref{G00})-(\ref{rhoi}) by using the new variable, $n = \ln a$ and normalize the $H^2$ and $H'$ by the present value of $F$ \ba && 3 F_{0} H^2  = 8 \pi G_{\ast} \rho_{\m} + \fr{1}{2} H^2 \phi^{'2} - 3 H^2 F' + 3 H^2 (F_0 - F) + U  \, , \label{G00n} \\ && 2 F_0 H H' = - 8 \pi G_{\ast} \rho_{\m} - H^2 \phi^{'2} - H^2 F'' + (H^2 - HH') F' + 2 HH' (F_0 - F) \, , \label{Giin} \\ && \phi'' + \Biggl( 3 + \fr{H'}{H} \Biggr) \phi' = 3 \Biggl( 2 + \fr{H'}{H} \Biggr) \fr{F'}{\phi'} - \fr{1}{H^2} \fr{U'}{\phi'} \, , \label{phieqn} \\ && \rho_{\m}^{'} + 3 \rho_{\m} = 0 \, , \label{rhoin} \ea where primes denote the differentiation w.r.t $n$, $F_{0}$ means $F(n=0)$ and we only consider the matter component which is relevant to the late-time universe. We will use the above background equations (\ref{G00n}) - (\ref{rhoin}) in followings.

\section{Perturbations}
\setcounter{equation}{0}

In this section, we review the matter density perturbations of STG in the longitudinal gauge w.r.t the JF \cite{0009034,Mukhanov} \be ds^2 = -(1 + 2 \Phi) dt^2 + a^2(1 - 2 \Psi) d \vec{x}^2 \, . \label{LG} \ee Also the energy momentum tensor components of the matter are given by \be T_{0}^{0} = - (\rho_{\m} + \delta \rho_{\m}) \, , \hspace{0.2in} T_{i}^{0} = - \rho_{\m} v_{i} \label{Tm} \ee where $\rho_{m} \, (\delta \rho_{m})$ is the energy density (contrast) of the pressureless matter and $v_{i}$ is its velocity. The gauge invariant quantity is defined as $\delta_{\m} \equiv \fr{\delta \rho_{\m}}{\rho_{\m}} + 3 H v$. The perturbed part of the energy momentum conservation in the Fourier space gives \ba \delta_{\m}' &=& -\fr{k^2}{a^2} \fr{v}{H} + 3 (\Psi + Hv)' \, , \label{dotdelta} \\ v' &=& \fr{\Phi}{H} \, , \label{dotv} \ea where $k$ is a comoving wavenumber. 
On the other hand, the first order perturbed Einstein equations give
\ba 3 F' \Phi' + \Biggl( 2 \lambda^{-2} F - Z \phi'^2 + 3 F' \Biggr) \Phi &=& - \Biggl[\fr{8 \pi G_{\ast} \rho_{\m}}{H^2} \delta_{\m} + \Biggl( \lambda^{-2} - 6 - 3 \fr{F'^2}{F^2} \Biggr) \delta F + \fr{\delta U}{H^2} \nonumber \\ && + 3 \fr{F'}{F} \delta F' + Z \phi' \delta \phi' + 3 Z \phi' \delta \phi + \fr{1}{2} \delta Z \phi'^2 \Biggr] \, , \label{dG00} \\ 2 F ( \Psi' + \Phi ) + F' \Phi &=& 8 \pi G_{\ast} \rho_{\m} \fr{v}{H} + Z \phi' \delta \phi + \delta F' - \delta F \, , \label{dG0i} \\ \Psi - \Phi &=& \fr{\delta F}{F} \, ,\label{dGij} \ea where $\lambda^2 = \fr{a^2H^2}{k^2}$. Also the perturbed part of the scalar field equation gives
\ba && \delta \phi'' + \Bigl(3 + \fr{H'}{H} + \fr{Z_{,\phi}}{Z} \phi' \Bigr) \delta \phi' + \Biggl[\lambda^{-2} -3(2 + \fr{H'}{H}) \Bigl( \fr{F_{,\phi}}{Z} \Bigr)_{,\phi} + \fr{1}{H^2} \Bigl( \fr{U_{,\phi}}{Z} \Bigr)_{,\phi} + \Bigl( \fr{Z_{,\phi}}{Z} \Bigr)_{,\phi} \fr{\phi'^2}{2} \Biggr] \delta \phi \nonumber \\ && = \Biggl[\lambda^{-2}(\Phi - 2\Psi) - 3 \Bigl( \Psi'' + (4 + \fr{H'}{H}) \Psi' + \Phi' \Bigr) \Biggr] \fr{F_{,\phi}}{Z} + ( 3\Psi' + \Phi' ) \phi' - 2 \fr{\Phi}{Z} \fr{U_{,\phi}}{H^2} \, . \label{deltaphieq} \ea
From the equations (\ref{dotdelta}) and (\ref{dotv}), we obtain \be \delta_{\m}'' + \Bigl( 2 + \fr{H'}{H} \Bigr) \delta_{\m}' + \lambda^{-2} \Phi = 3 (\Psi + H v)'' + \Bigl( 6 + 3 \fr{H'}{H} \Bigr) (\Psi + H v)' \, . \label{ddotdelta} \ee
We neglect time derivative terms w.r.t spatial derivative terms of corresponding perturbed variables ({\it i.e.} subhorizon limit). This simplification holds at scales $k \ga aH \la 10^{-3} h/$Mpc. In this limit we can write the approximate equations for (\ref{dG00}) and (\ref{deltaphieq}) \ba \lambda^{-2} \Phi &\simeq& - \fr{1}{2F} \Biggl[\fr{8 \pi G_{\ast} \rho_{\m}}{H^2} \delta_{\m} + \lambda^{-2} \delta F \Biggr] \, , \label{dG00app} \\ \delta \phi &\simeq& (\Phi - 2 \Psi) \fr{F_{,\phi}}{Z} \simeq - \fr{F F_{,\phi}}{Z F + 2 F_{\, ,\phi}^2} \Phi \, , \label{deltaphiapp} \ea where we use Eq. (\ref{dGij}) in the second equality of Eq. (\ref{deltaphiapp}). From the above two equations, we have the Poisson's equation \be \fr{k^2}{a^2} \Phi \simeq - 4 \pi G_{\eff} \rho_{\m} \delta_{\m} \, , \label{Poisson} \ee where $G_{\eff}$ is given in (\ref{Geff}). Combining equations (\ref{ddotdelta}) and (\ref{Poisson}) gives \be \delta_{\m}'' + \Bigl( 2 + \fr{H'}{H} \Bigr) \delta_{\m}' - \fr{4 \pi G_{\eff} \rho_{\m}}{H^2} \delta_{\m} \simeq 0 \, . \label{ddotdelta2} \ee We can find the anisotropic parameter $\eta$ defined by $\eta = (\Phi - \Psi)/\Psi$ \cite{07042421} from the equations (\ref{dGij}) and (\ref{deltaphiapp}) \be \eta \simeq - \fr{F_{,\phi}^2}{Z F + 3 F_{,\phi}^2} \, . \label{eta} \ee

\section{Constraints of $F(\phi)$ and $U(\phi)$}
\setcounter{equation}{0}

From the observational viewpoint, it is quite useful to use a parametrization of both the expansion history and the growth factor of the matter perturbation in terms of the EOS $\omega(n)$ and the GIP $\gamma(n)$. Chevallier-Polarski-Linder (CPL) parameterization $\omega = \omega_{0} + \omega_{a} (1 - e^{n})$ uses its present value ($\omega_{0}$) and variation ($\omega_{a}$) and it might be suitable for future observations \cite{0009008,0208512}. The growth factor can be given by $\delta_{\m}(n) = e^{n} g(n)$ where $g(n)$ is given in Eq. (\ref{Linderg}) with $\gamma = 0.55 + 0.05 (1 + \omde (z=1))$\cite{0507263,9804015}. Although this functional form is accurate at certain level, there are several drawbacks in it. We explain the details about these problems in the appendix.

Eq. (\ref{ddotdelta2}) can be rewritten if we use the definition of the GIP $\gamma$ by using $f = \fr{d \ln \delta_{\m}(n)}{d n} \equiv \Omn^{\gamma}$ \cite{Peebles1,Peebles2,Lahav} \ba && \Omn^{2 \gamma} + \Biggl( \gamma' \ln \Omn + \gamma \fr{\Omn'}{\Omn} + 2 + \fr{H'}{H} \Biggr) \Omn^{\gamma} -\fr{3}{2} \fr{F_{0}}{F} \Biggl[ \fr{2F + 4 (F'/\phi')^2}{2F + 3 (F'/\phi')^2} \Biggr] \Omn \nonumber \\ && \simeq  \Omn^{2 \gamma} + \Biggl( \gamma' \ln \Omn + \gamma \fr{\Omn'}{\Omn} + 2 + \fr{H'}{H} \Biggr) \Omn^{\gamma} -\fr{3}{2} \fr{F_{0}}{F} \Omn = 0 \, , \label{geq} \ea where we use the assumption $|F(n)| \gg |F_{, \, \phi}(n)^2|$ in the approximation. Thus, $F(n)$ is given by \ba \fr{F(n)}{F_{0}} &=& \fr{3}{2} \fr{\Omn}{P(n)} \,\,\,\, , {\rm where} \nonumber \\ P(n) &=& \Omn^{\gamma} \Biggl( \Omn^{\gamma} + \gamma' \ln \Omn + \gamma \fr{\Omn'}{\Omn} + 2 + \fr{H'}{H} \Biggr) \label{FoF0} \\ &=& \Omn^{\gamma} \Biggl( \Omn^{\gamma} + \gamma' \ln \Omn - \gamma \Bigl[ 3 + 2 \fr{H'}{H} \Bigr] + 2 + \fr{H'}{H} \Biggr) \, , \nonumber \ea where we use the relation $\Omn' = - (3 +2 H'/ H) \Omn$ in the second equality of $P(n)$. When we adopt CPL $\omega$, then $H^2$ and $H'/H$ are given by
\ba \fr{H^2}{H_0^2} &=& \Omo e^{-3n} + (1 - \Omo) e^{-3(1 + \oo + \oa) n} e^{-3 \oa (1 - e^{n})} \, , \label{H2oH0} \\
\fr{H'}{H} &=& -\fr{3}{2} \Biggl[ 1 + \omega \fr{(1 - \Omo) e^{-3(\oo + \oa)n} e^{-3 \oa (1 - e^{n})}}{\Omo + (1 - \Omo) e^{-3(\oo + \oa)n} e^{-3 \oa (1 - e^{n})}} \Biggr] \, . \label{HpoH} \ea
$F_{0} \simeq 1$ in order to satisfy $G_{\eff}(0) \simeq G_{N}$. After we obtain the functional form of $F(n)$, we can also derive the that of the potential $U(n)$ by using Eqs. (\ref{G00n}) and (\ref{Giin}) \be \fr{U(n)}{F_{0} H_{0}^2} = \fr{1}{2} \fr{H^2}{H_{0}^2} \Biggl( \fr{F''}{F_0} + \Bigl[ 5 + \fr{H'}{H} \Bigr] \fr{F'}{F_0} + 2 \Bigl[ 3 + \fr{H'}{H} \Bigr] \fr{F}{F_0} - 3 \Omn \Biggr) \, . \label{U} \ee The effective equation of state $\omega$ is obtained from the same equations \be \omega = \fr{\phi'^2 + 2 F'' +  2 (2 + H'/H) F' + 4 (F - F_{0}) H'/H + 6 (F - F_{0}) - 2U/H^{2}}{\phi'^2 - 6F' - 6(F - F_{0}) + 2U/H^{2}} \, , \label{omde} \ee where the evolution equation of scalar field $\phi'$ is given from Eq. (\ref{Giin}) \be \phi' = \sqrt{-F'' + \Bigl(1 - \fr{H'}{H} \Bigr) F' - 2\fr{H'}{H} F - 3 F_{0} \Omega_{\rm{m}}} \, . \label{phiprime} \ee Thus, the EOS obtained from Eq. (\ref{omde}) should be same as that of CPL. As long as $F_{0} \simeq 1$ is satisfied, the above constraint is well matched. We obtain the useful recursion relations for the differentiation of $F$ from Eq. (\ref{FoF0}) \ba \fr{F'}{F_0} &=& - \Biggl( 3 + 2 \fr{H'}{H} + \fr{P'}{P} \Biggr) \fr{F}{F_{0}} \, , \label{FpoF0} \\ \fr{F''}{F_0} &=& \Biggl( \Bigl[3 + 2 \fr{H'}{H} + \fr{P'}{P} \Bigr]^2 - 2 \Bigl[ \fr{H'}{H} \Bigr]' - \Bigl[ \fr{P'}{P} \Bigr]' \Biggr) \fr{F}{F_0} \, . \label{FppoF0} \ea

Now we probe the viabilities of some specific models which can mimic some well known dark energy and modified gravity models. We assume that $\gamma = \gamma_0 + \gamma_a (1 - a)$ and $\Omo = 0.3$ in the following analysis. One can always find the constant value of $\gamma$ ({\it i.e.} $\gamma_a = 0$) from Eq. (\ref{FoF0}) by using $F(0)/F_0 = 1$.

\subsection{Comparison with Dark Energy models}
\begin{figure}
\centering
\vspace{1.5cm}
\begin{tabular}{ccc}
\epsfig{file=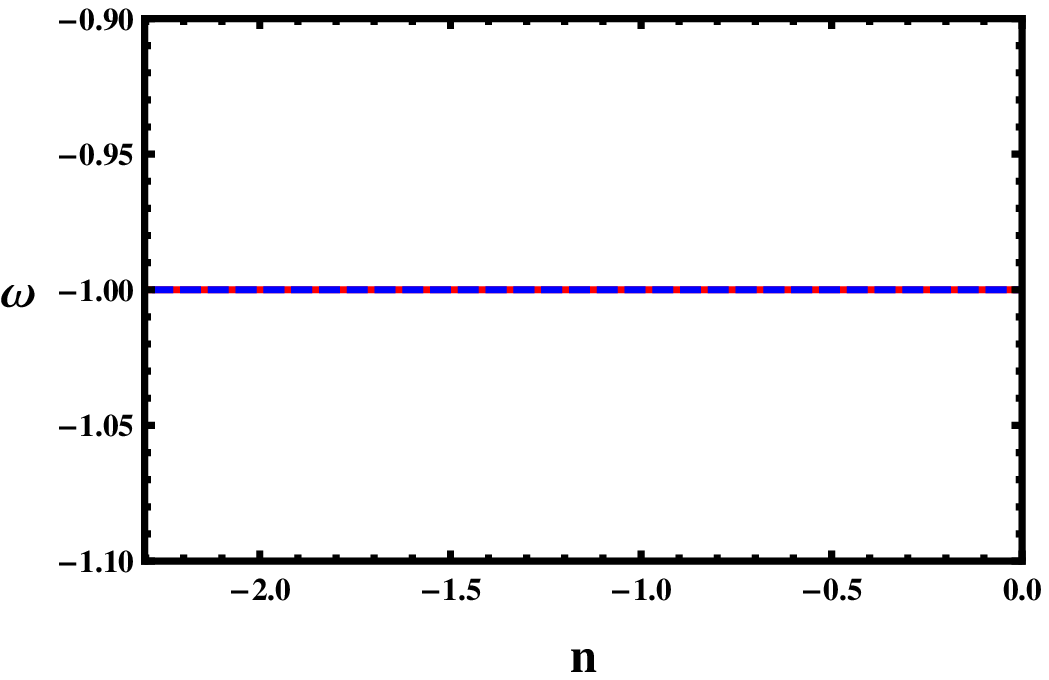,width=0.3\linewidth,clip=} &
\epsfig{file=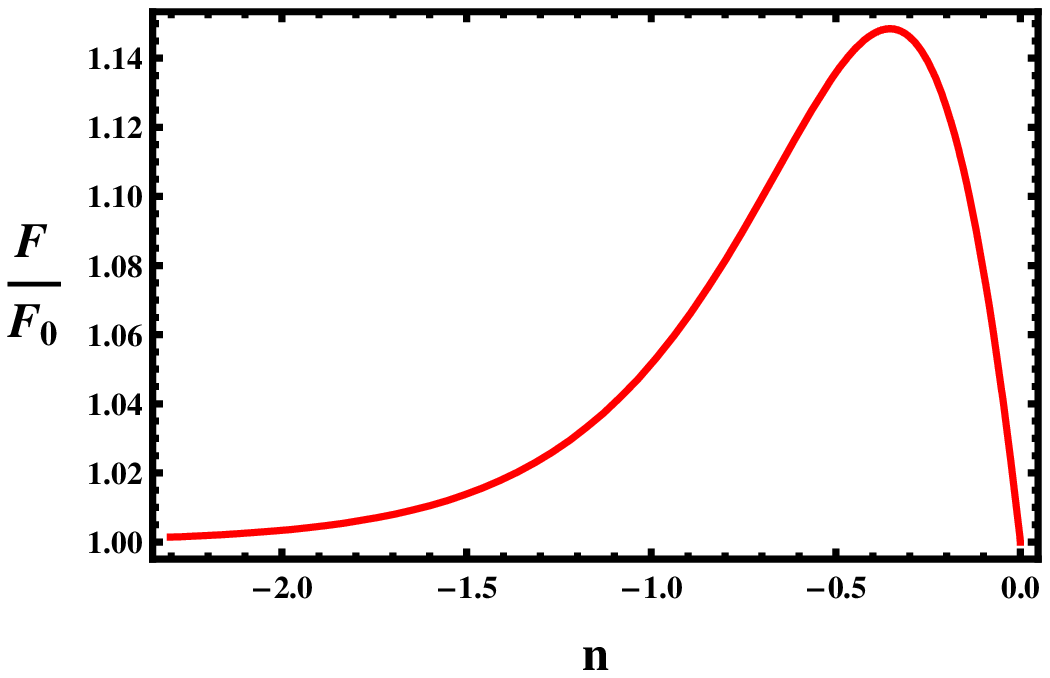,width=0.3\linewidth,clip=} &
\epsfig{file=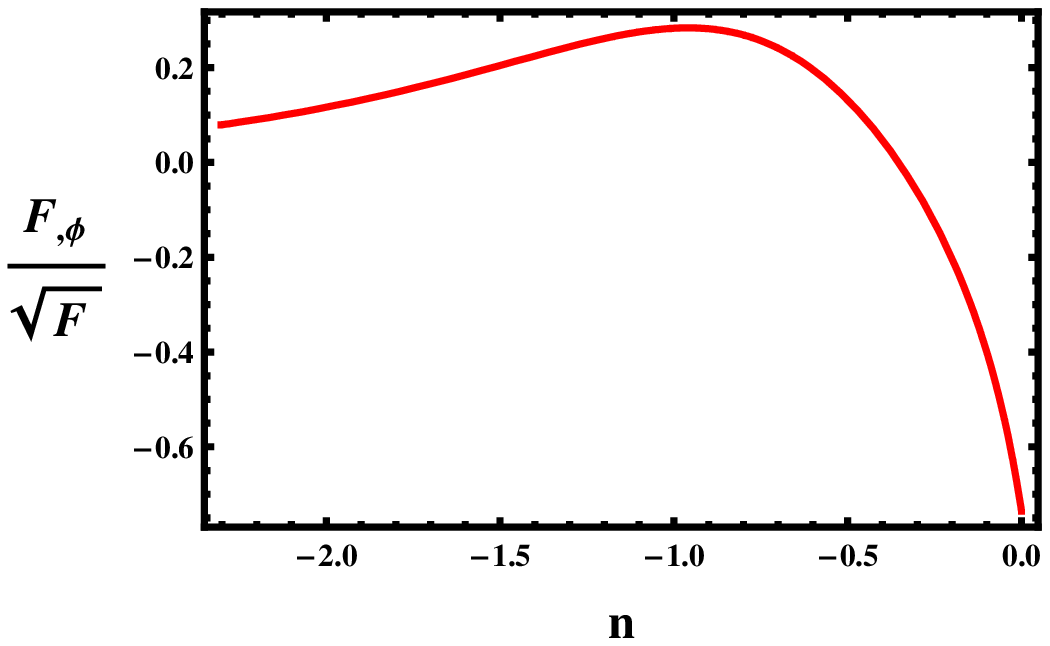,width=0.31\linewidth,clip=} \\
\epsfig{file=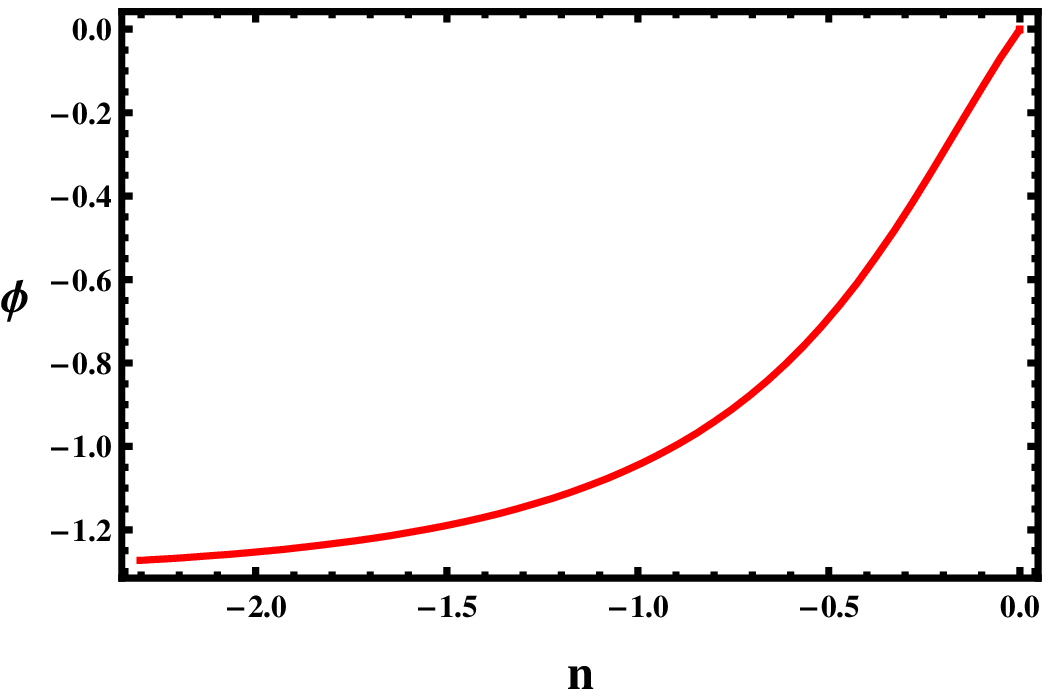,width=0.3\linewidth,clip=} &
\epsfig{file=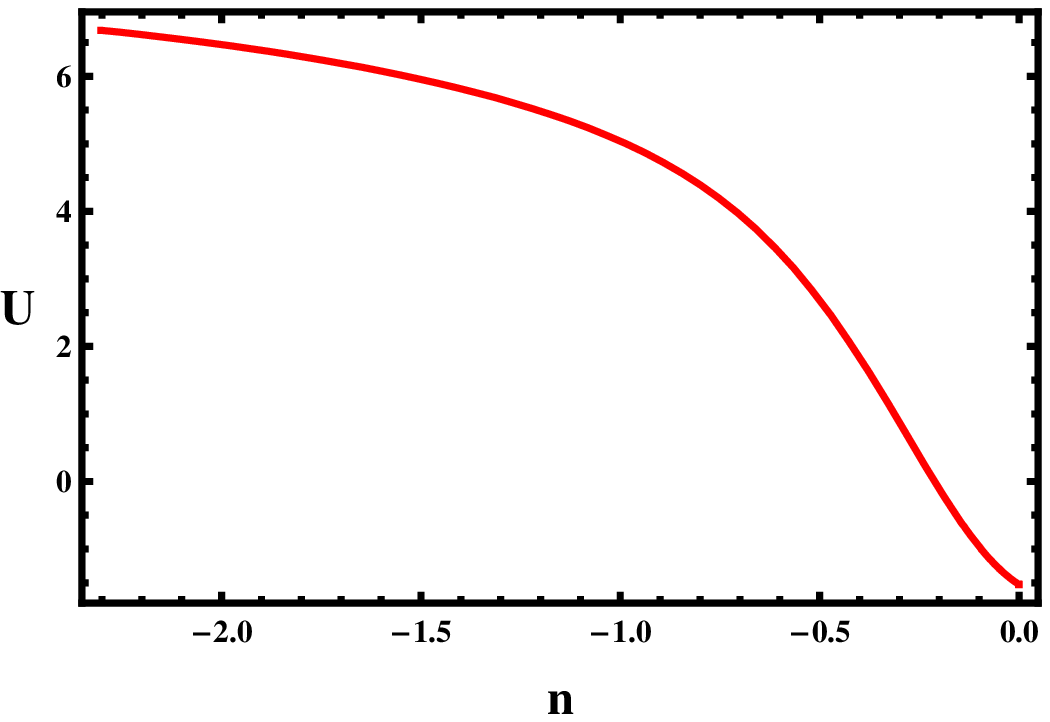,width=0.29\linewidth,clip=} &
\epsfig{file=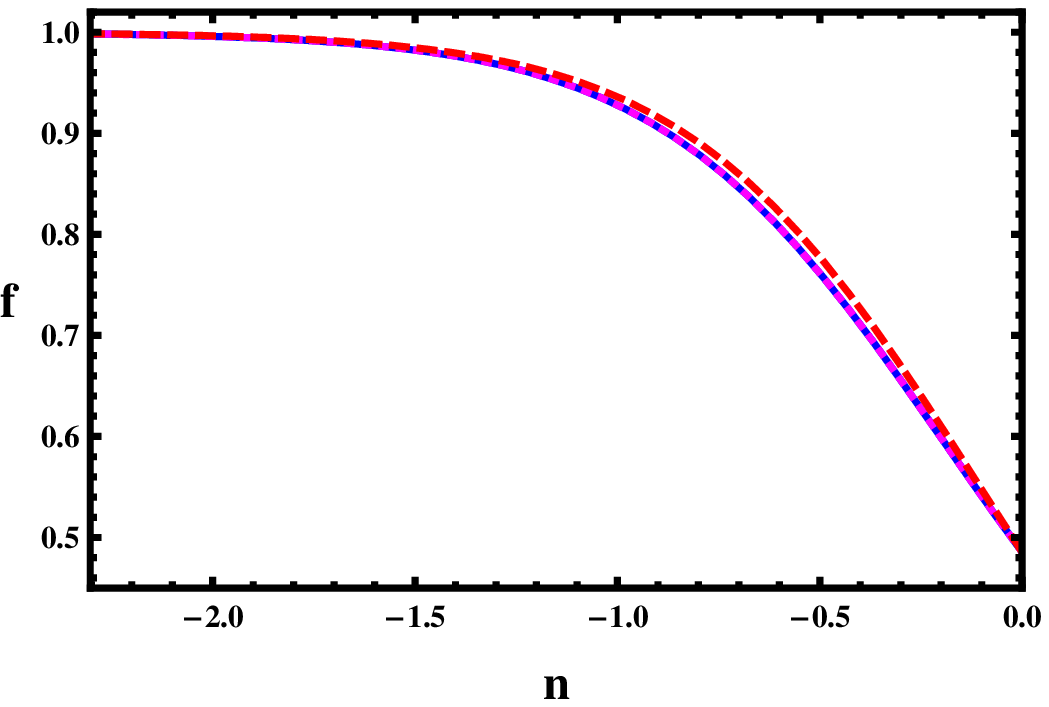,width=0.3\linewidth,clip=} \\
\end{tabular}
\vspace{-0.5cm}
\caption{ a) In the first row, we show the evolutions of $\omega$, $F/F_0$, and $F_{\, , \phi} / \sqrt{F}$ (from left to right) when ($-1.0, 0$) and ($0.6, 0.126$) for ($\oo, \oa$) and ($\gamma_0, \gamma_a$), respectively.  b) The evolutions of $\phi$, $U$, and $f$ for the same values of $\omega$ and $\gamma$ (from left to right) in the second row.} \label{fig1}
\end{figure}

First, we investigate the models which mimic the cosmological constant ($\oo, \oa$) = ($-1.0, 0$) in the background evolution. We can further specify the models by considering the different growth history ({\it i.e.} different value of $\gamma$). When $\gamma_{a} = 0$, $\gamma_{0} \simeq 0.56$ is found to get $F(0) = 1$. However, this gives $(F_{\, , \phi}/\sqrt{F}) |_{0} = -0.2$ which violates the solar system test. The parameter set ($\gamma_{0}, \gamma_{a}$) = ($0.56, -0.012$) gives $F(0) = 1.017$ and $(F_{\, , \phi}/\sqrt{F}) |_{0} = -0.019$ and passes the solar system test. But $\phi'$ becomes imaginary at $n < -0.6$ ({\it i.e.} $z \ga 0.8$) in this case. This is quite similar to the result in Ref. \cite{0009034} where $F(n)/F_0$ is defined as $e^{-2n} f(n)$ with $f(n)$ is a general function of $n$. Thus, we may need to consider general $Z(\phi)$ in order to avoid this singular behavior of $\phi'$ \cite{SLeefu}. Even though the above case ($0.56, -0.012$) might be cured by introducing the nontrivial $Z(\phi)$, it is difficult to be distinguished from the cosmological constant ($\Lambda$) model because the values of both $\omega$ and $\gamma$ are quite similar to those of $\Lambda$ model. Thus, it is worth to check the case when $\gamma$ is quite different from that of $\Lambda$ model. We try with ($\gamma_{0}, \gamma_{a}$) $\simeq$ ($0.6, 0.126$) which gives the quite interesting evolutions of physical quantities. Even though the potential $U(n)$ is negative at present, it is fine because the second derivative of $U(\phi)$ does not give the precise value of its squared mass in JF. We show the evolutions of physical quantities in Fig. \ref{fig1}. As shown in the first column at the first row, $\omega$ obtained from Eq. (\ref{omde}) is exactly same as that of CPL with $\oo = -1.0$ and $\oa = 0$. Thus, if one investigate only the geometrical tests, then this STG theory can not be distinguished from $\Lambda$ model. However, $F(n)/F_0$ can reach to $1.5$ around $n \sim -0.4 - -0.3$ ({\it i.e.} $z \sim 0.3 - 0.4$), which can give some significant effects on ISW or WL. The growth index $f$ obtained from general $\gamma$ (dotted line) is also consistent with the exact solution (solid one). The dashed line in $f$ deviating from the exact solution is the one obtained when one use the constant $\gamma = \gamma_0 = 0.6$. The only problem in this model is that it violates the solar system limit by $F_{\, , \phi} / \sqrt{F} |_{0} \simeq -0.7$. If one release this limit which might be plausible in cosmological scale, then we may have a very interesting STG model which mimics exactly $\Lambda$ model for the background evolution but shows the totally different behavior for the growth history. When we consider $\gamma_0 < 0.56$, the value of $\phi'$ becomes imaginary at the present. Thus, STG with simple form of $Z(\phi) = 1$ may not be consistent with small value of $\gamma_0$ ({\it i.e.} negative $\gamma_a$). The shape of $F(n)/F_0$ is determined by the sign of $\gamma_a$. If $\gamma_a$ is positive (negative), then $F(n)/F_0$ shows the convex (concave) shape with the minimum (maximum) as $1$. Thus, the value of $F(n)$ is bigger than $F_0$ in the past for the viable STG.

\begin{figure}
\centering
\vspace{1.5cm}
\begin{tabular}{ccc}
\epsfig{file=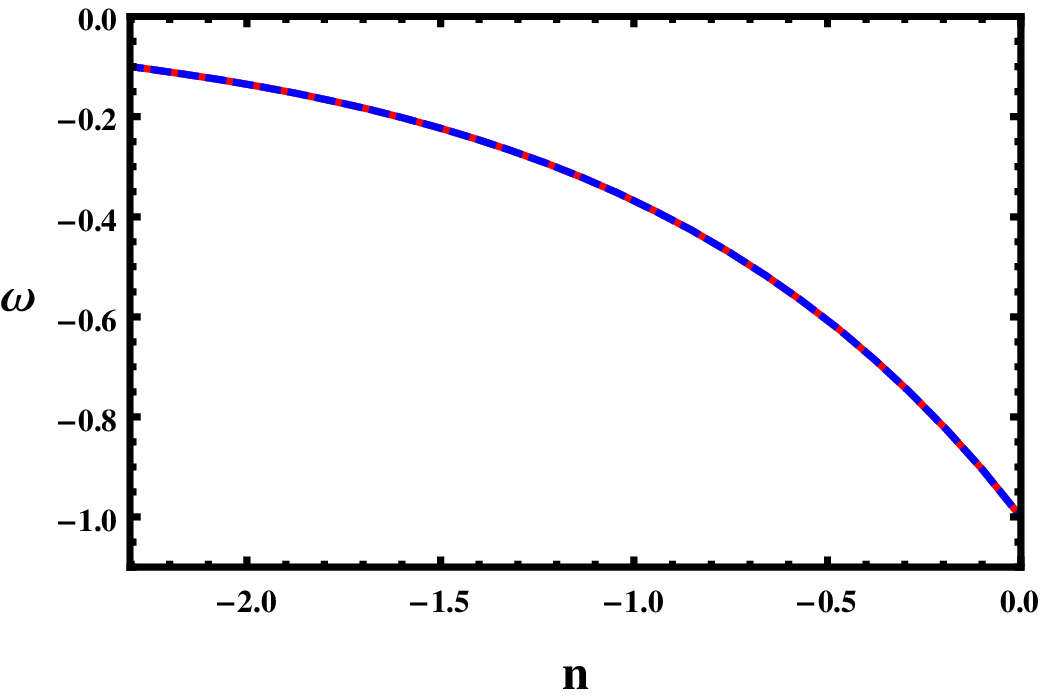,width=0.3\linewidth,clip=} &
\epsfig{file=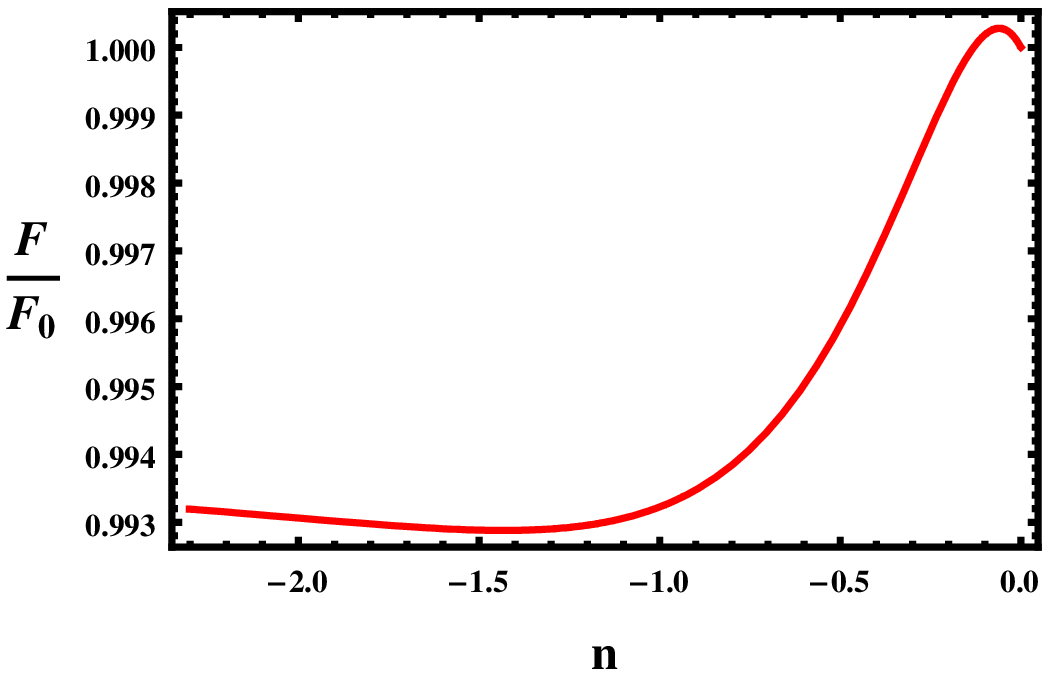,width=0.3\linewidth,clip=} &
\epsfig{file=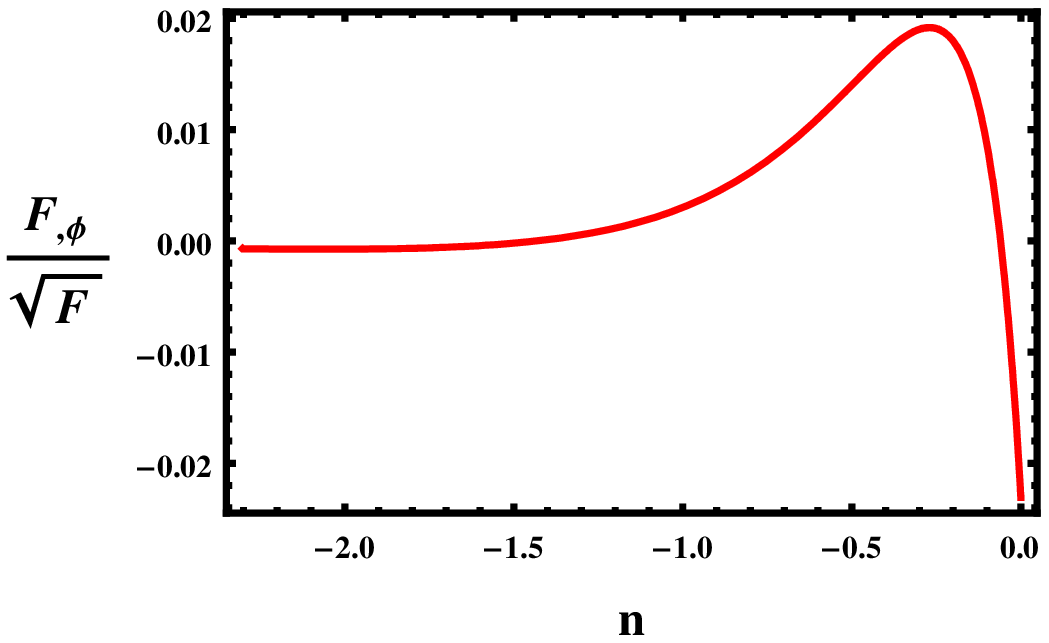,width=0.31\linewidth,clip=} \\
\epsfig{file=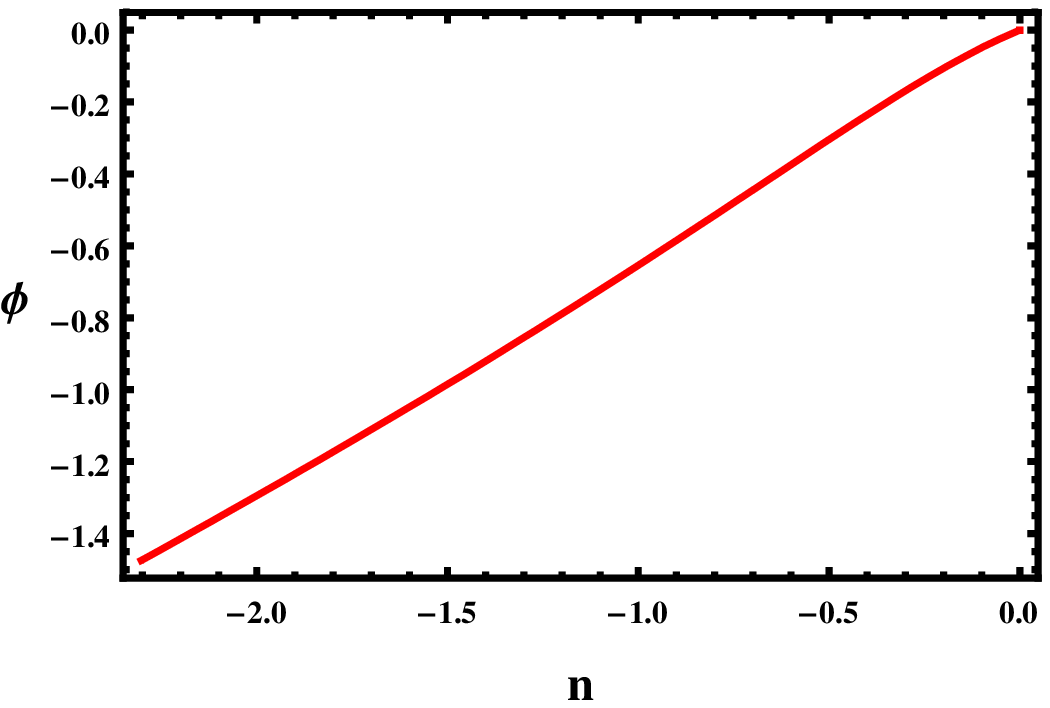,width=0.3\linewidth,clip=} &
\epsfig{file=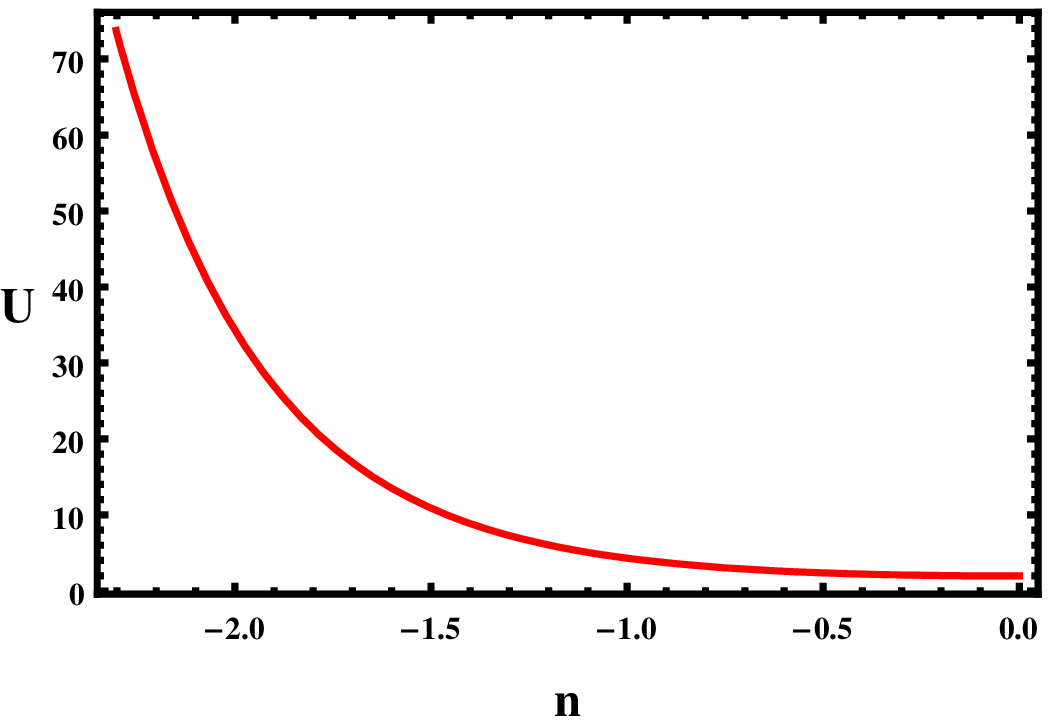,width=0.29\linewidth,clip=} &
\epsfig{file=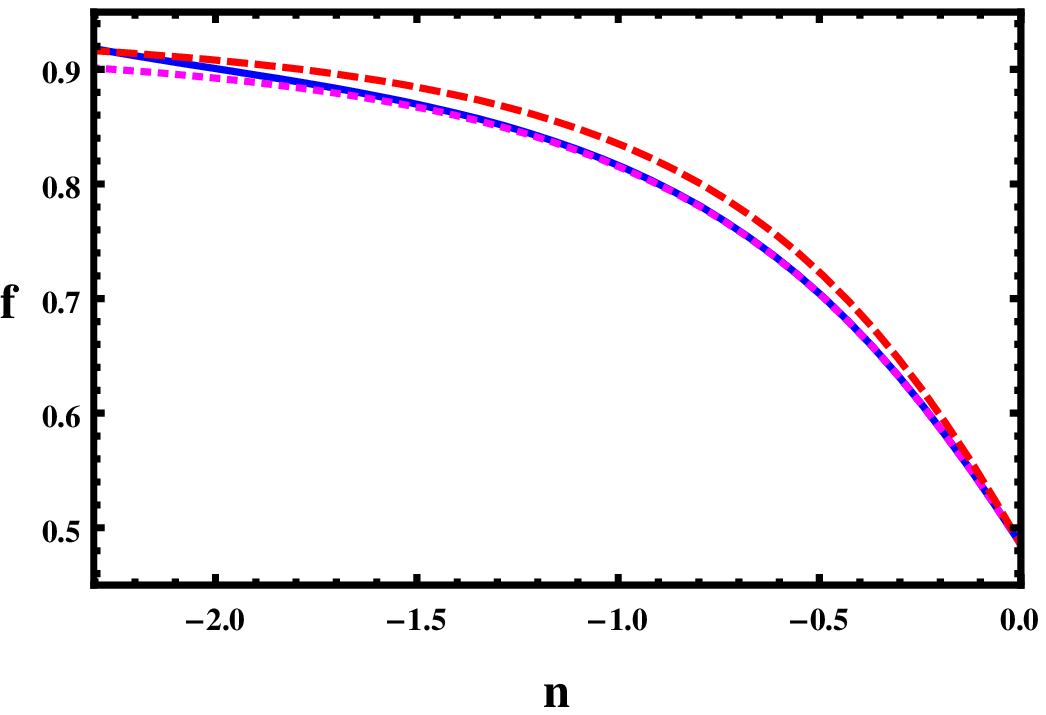,width=0.3\linewidth,clip=} \\
\end{tabular}
\vspace{-0.5cm}
\caption{ a) In the first row, we show the evolutions of $\omega$, $F/F_0$, and $F_{\, , \phi} / \sqrt{F}$ (from left to right) when ($-1.0, 1.0$) and ($0.562, 0.006$) for ($\oo, \oa$) and ($\gamma_0, \gamma_a$), respectively.  b) The evolutions of $\phi$, $U$, and $f$ for the same values of $\omega$ and $\gamma$ (from left to right) in the second row.} \label{fig2}
\end{figure}

Second, we probe the models which show the similar background evolutions as the some coupled quintessence models ($\oo, \oa$) = ($-1.0, 1.0$). In these models, $\omega = 0$ during the tracking region and it approaches to $-1$ at present \cite{9904120,0601333}. Again we find $\gamma_{0} \simeq 0.56$ for $\gamma_{a} = 0$ to get $F(0) = 1$. However, these values of STG model can not pass the solar system test because $(F_{\, , \phi}/\sqrt{F}) |_{0} = -0.17$. STG with other set of parameters ($\gamma_{0}, \gamma_{a}$) $\simeq$ ($0.562, 0.006$) gives $F(0) = 1.00$ and $(F_{\, , \phi}/\sqrt{F}) |_{0} = -0.02$. Also $\phi'$ well behaves in this case. Thus, this model satisfy all known observational constraints. We show the evolutions of physical quantities of this in Fig. \ref{fig2}. Again $\omega$ obtained from Eq. (\ref{omde}) is exactly same as CPL $\omega$ with $\oo = -1.0$ and $\oa = 1.0$. $F(n)$ decreases during the evolution which is opposite to the previous case even though $\gamma_a > 0$. However, this is only possible when $\gamma \simeq 0$. The growth index $f$ obtained from general $\gamma$ (dotted line) is also consistent with the exact solution (solid line). Again $f$ with constant value of $\gamma_0 = 0.562$ is deviated from the exact one. For ($\gamma_{0}, \gamma_{a}$) $\simeq$ ($0.6, 0.126$),  $F(n)/F_0$ increases to $1.1$ around $n \sim -0.4$ ({\it i.e.} $z \sim 0.4$). This case also violates the solar system limit by $F_{\, , \phi} / \sqrt{F} |_{0} \simeq -0.45$.

In Table \ref{table1}, we show the behaviors of physical quantities for various models with some parameter sets. There have been various quintessence models \cite{0604602} but we just investigate the inverse power-law potentials ($\phi^{\alpha}$) \cite{Ratra,9809272} with $\alpha = 1$ which is well approximated by $\oo = -0.74$ and $\oa = 0.07$. STG for the same values of $\oo$ and $\oa$ with $\gamma_0 = 0.57$ is hardly distinguished from the $\phi^{-1}$ quintessence model because $f$ of both models are quite similar to each other. $F(n)/F_0$ reaches to the maximum value $1.015$ and then approaches to $1$ at early time. No known observation can distinguish this from $\phi^{-1}$ quintessence model. If ($\gamma_0, \gamma_a$) are ($0.6, 0.08$) for the same $\omega$, then the maximum value of $F(n)/F_0$ is $1.15$ which can give some effects on ISW and WL. Also the time varying growth index parameter can be distinguished from the constant one. We also investigate the model inspired by the supergravity (SUGRA) \cite{0208156}. The background evolution of SUGRA can be parameterized by ($\oo, \oa$) $=$ ($-0.92, -0.08$). Again, only STG with $\gamma_0 = 0.6$ with $\gamma_a = 0.11$ can give some significant deviations from SUGRA for the growth history. We also check the phantom crossing models in this table. Phantom crossing I means the model with $\omega > -1$ in the past becomes $\omega < -1$ at present. Phantom crossing II indicates the other model where $\omega < -1$ in the past and $\omega > -1$ at present. The results do not change much for these cases. From this table, we find that only STG models with the high values of $\gamma_0$ can be distinguished from the other models. However, all of high values of $\gamma_0$ STG models violate the solar system test.

\begin{center}
    \begin{table}
    \begin{tabular}{ | c | c | c | c | c | c | c | c |  }
    \hline
      $\rm{Models}$ & $\oo$  & $\oa$ & $\gamma_0$ & $\gamma_a$& $F(n)/F_0$ & $F_{\, , \phi}/\sqrt{F} |_0$ & $\phi'$ \\ \hline
      $ $ & $$  & $$ & $0.56$ & $-0.018$& $\bigcup$ \rm{min} $0.986$ & $0.018$ & \rm{fine}  \\ \cline{4-8}
      $V(\phi) \propto \phi^{-1}$ & $-0.74$  & $0.07$ & $0.57$ & $0$& $\bigcap$ \rm{max} $1.015$ & $-0.09$ & \rm{fine}  \\ \cline{4-8}
      $ $ & $ $  & $ $ & $0.6$ & $0.08$& $\bigcap$ \rm{max} $1.150$ & $-0.48$ & \rm{fine}  \\ \hline
      $ $ & $$  & $$ & $0.56$ & $-0.016 $& $\nearrow$ \rm{max} $1.01$ & $0.00$ & $z < 1.3$  \\ \cline{4-8}
      $\rm{SUGRA}$ & $-0.92$  & $-0.08$ & $0.563$ & $0$& $\bigcap$ \rm{max} $1.02$ & $-0.18$ & \rm{fine}  \\ \cline{4-8}
      $ $ & $ $  & $ $ & $0.6 $ & $0.11 $& $\bigcap$ \rm{max} $1.14$ & $-0.70$ & \rm{fine}  \\ \hline
      Phantom & $$  & $$ & $0.53$ & $-0.09 $& $\bigcup$ \rm{min} $0.93$ & $\rm{none}$ & imaginary  \\ \cline{4-8}
      Crossing & $-1.1$  & $0.3$ & $0.557$ & $0$& $\bigcap$ \rm{max} $1.01$ & $-0.24$ & \rm{fine}  \\ \cline{4-8}
      I & $ $  & $ $ & $0.6$ & $0.143$& $\bigcap$ \rm{max} $1.15$ & $-0.68$ & \rm{fine}  \\ \hline
      Phantom & $$  & $$ & $0.55$ & $-0.049 $& $\bigcup$ \rm{min} $0.97$ & none & imaginary  \\ \cline{4-8}
      Crossing & $-0.8$  & $-0.3$ & $0.568$ & $0$& $\bigcap$ \rm{max} $1.03$ & $-0.18$ & $z < 3$  \\ \cline{4-8}
      II & $ $  & $ $ & $0.6$ & $0.09 $& $\bigcap$ \rm{max} $1.13$ & $-0.69$ & $z < 7$  \\ \hline
    \end{tabular}
    \caption{Physical quantities for various models. $\bigcup$ ($\bigcap$) means the concave (convex) shape of $F(n)/F_0$. Also $\nearrow$ indicates the monotonic increase of $F/F_0$ as $n$ increases. $\phi'$ becomes imaginary after the maximum $z$ for certain parameter set. }
    \label{table1}
    \end{table}
\end{center}

\subsection{Comparison with DGP model}

\begin{figure}
\centering
\vspace{1.5cm}
\begin{tabular}{ccc}
\epsfig{file=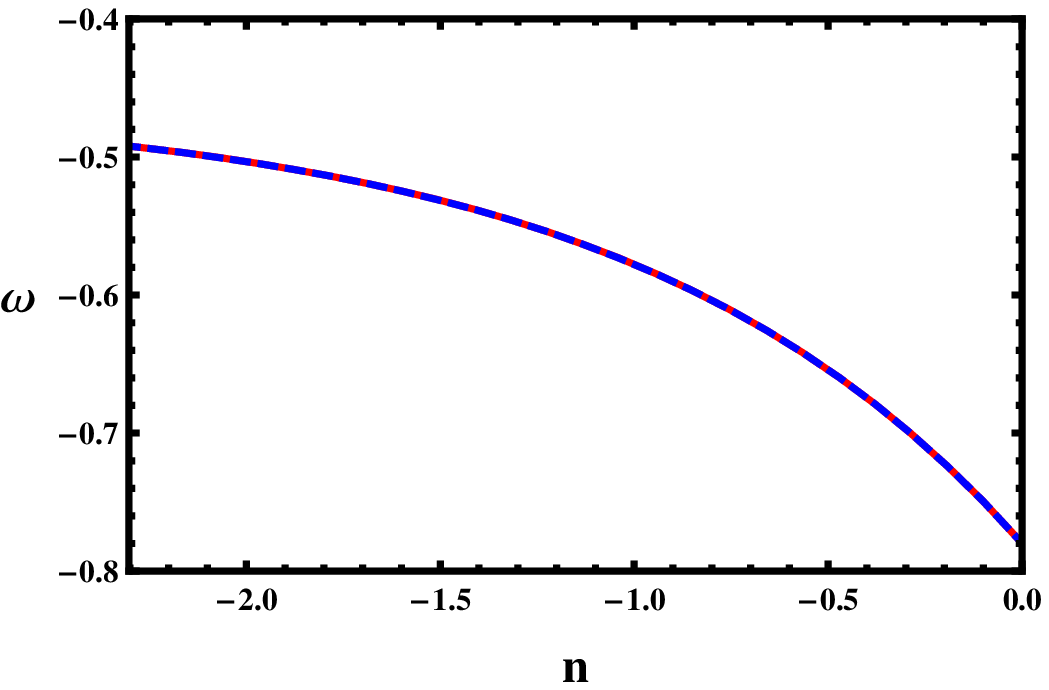,width=0.3\linewidth,clip=} &
\epsfig{file=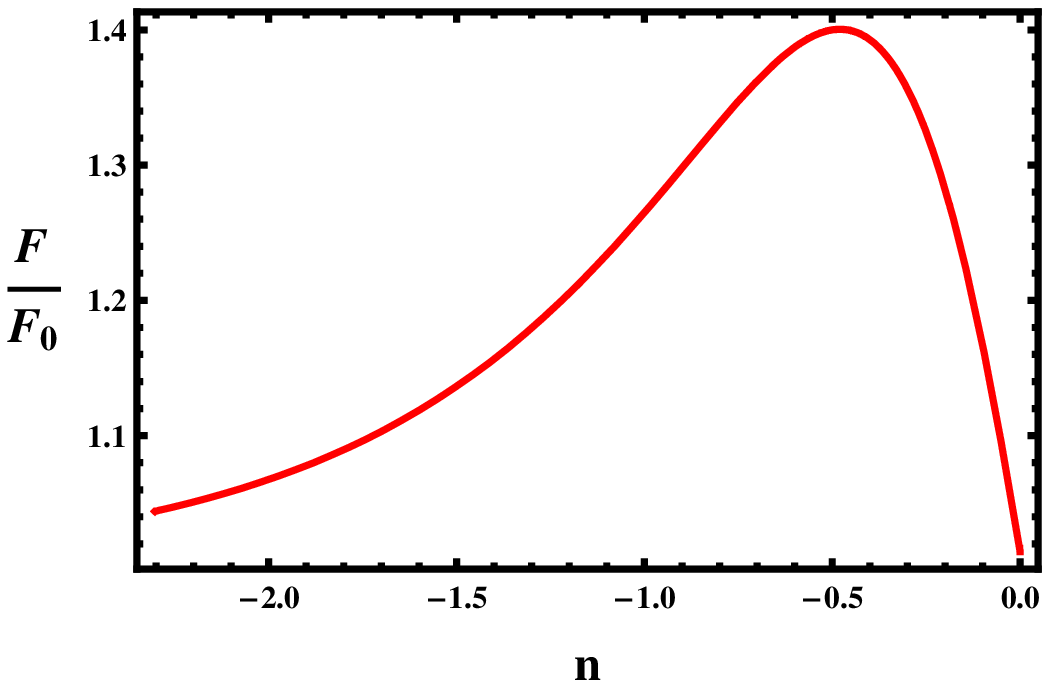,width=0.3\linewidth,clip=} &
\epsfig{file=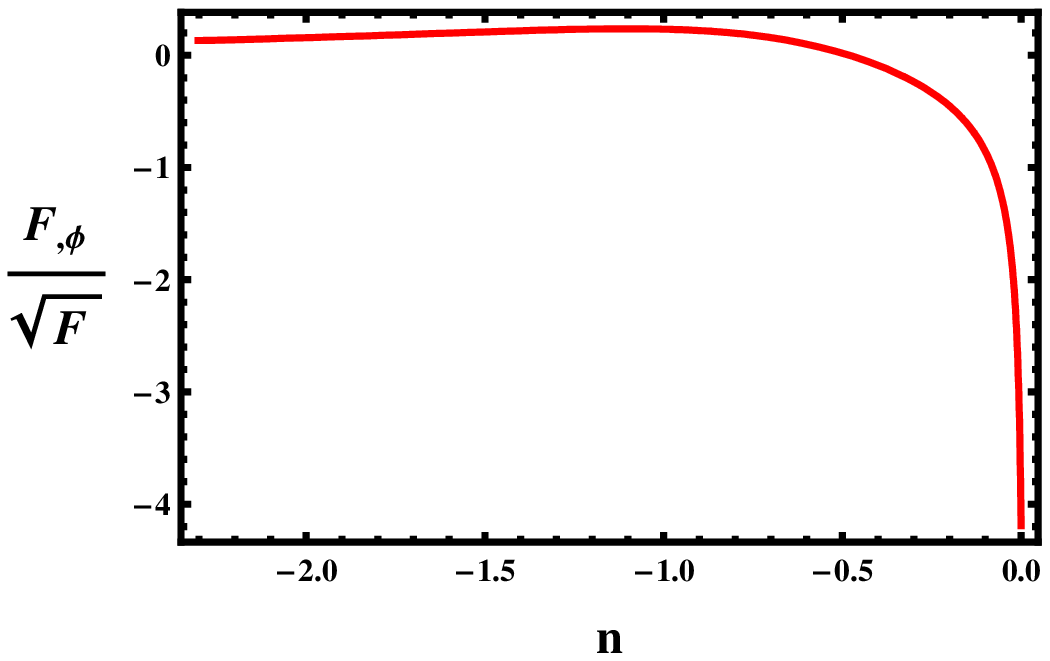,width=0.31\linewidth,clip=} \\
\epsfig{file=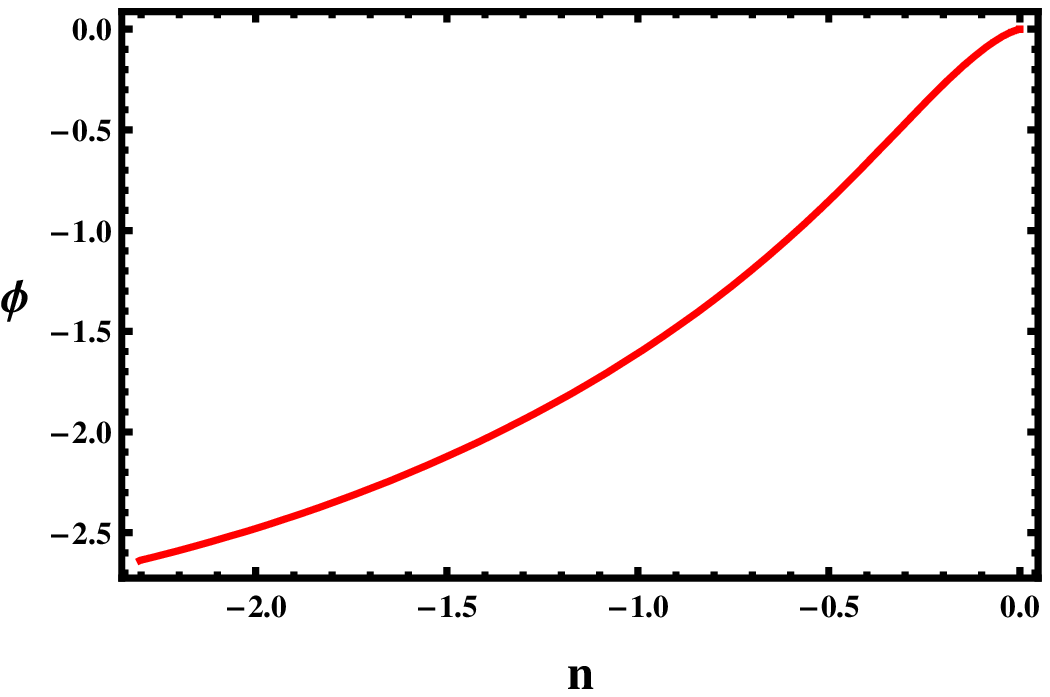,width=0.3\linewidth,clip=} &
\epsfig{file=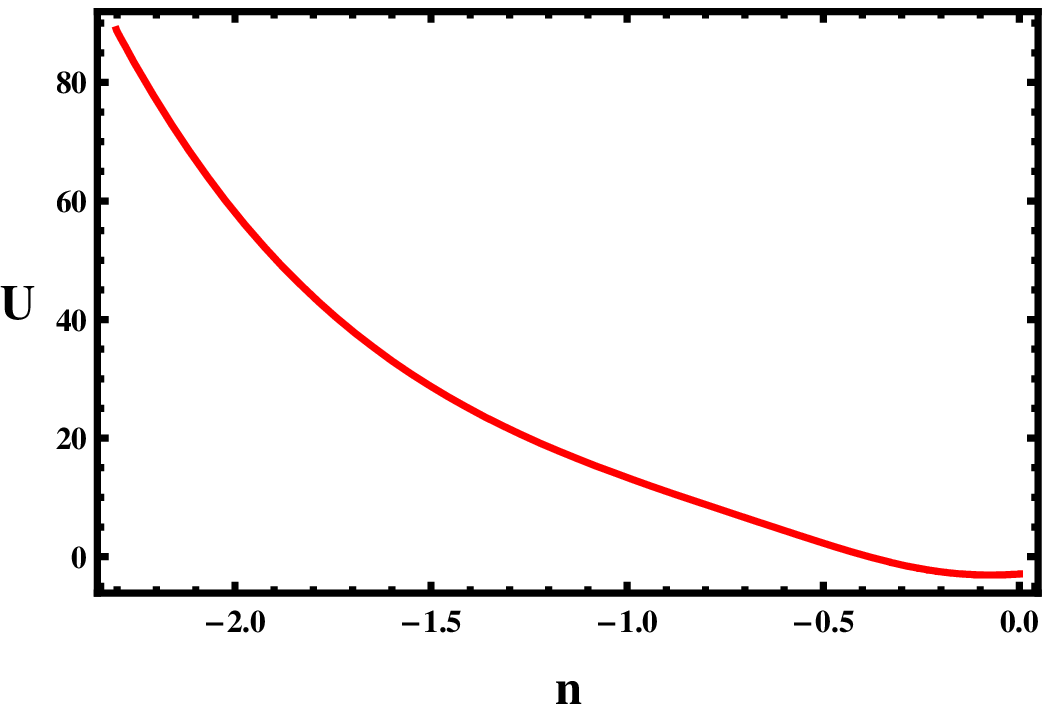,width=0.29\linewidth,clip=} &
\epsfig{file=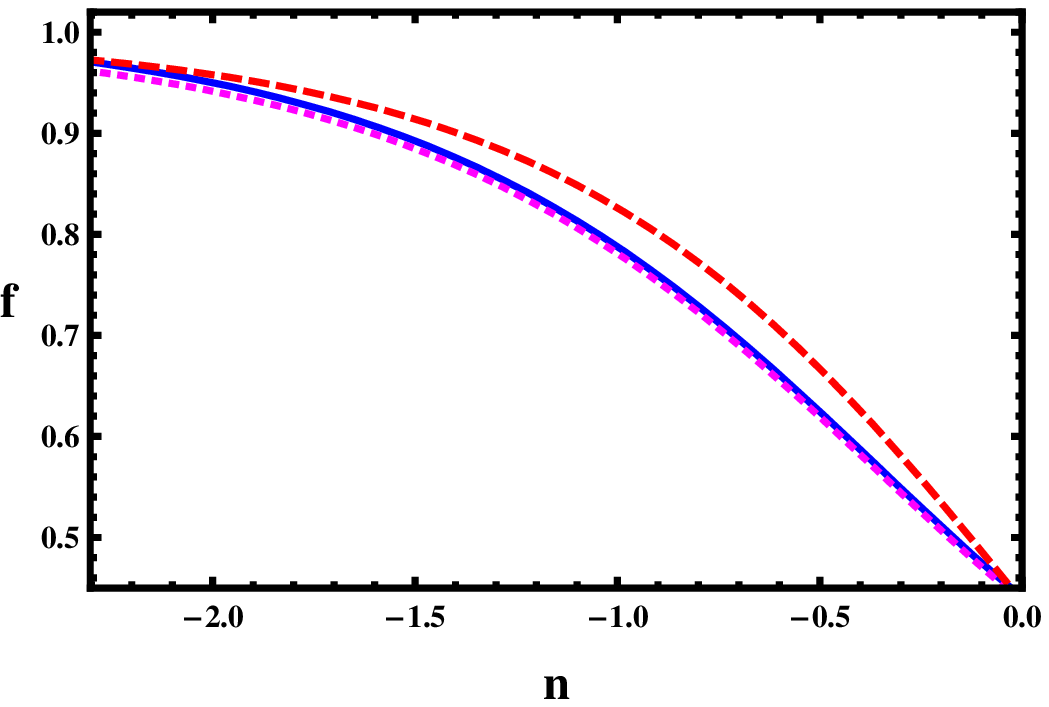,width=0.3\linewidth,clip=} \\
\end{tabular}
\vspace{-0.5cm}
\caption{ a) In the first row, we show the evolutions of $\omega$, $F/F_0$, and $F_{\, , \phi} / \sqrt{F}$ (from left to right) for STG model with ($\gamma_0, \gamma_a$) $=$ ($11/16, 0.32$).  b) The evolutions of $\phi$, $U$, and $f$ for the same values of $\omega$ and $\gamma$ (from left to right) in the second row.} \label{fig3}
\end{figure}

DGP model \cite{0005016,0010186,0105068} is one of the well studied modified gravity models. The background evolution of this model can be parameterized by ($\oo, \oa$) $=$ ($-0.78, -0.32$) with $\gamma_0 = 11/16$ \cite{0507263}. In Ref. \cite{08024122}, $\gamma_0 = 4/7$ at high $z$. We show the evolutions of physical quantities of STG with ($\gamma_0,\gamma_a$) $=$ ($11/16, 0.32$) in Fig. \ref{fig3}. Again except the solar system constraints, this model shows the interesting and different behaviors compared to DGP model ($\gamma_a = 0$) for the growth history. $F_{\, , \phi}/F |_0$ becomes very big negative value and this is due to the fact that $\phi'$ approaches to $0$ at present which is used in $F_{\, , \phi} = F' / \phi'$. Usually, $f$ is smaller than $1$ in general DE and MG models. However, we can produce $f > 1$ for $\gamma_0 < 0.5$ with non-vanishing $\gamma_a$ models even though $\phi'$ is imaginary. If observations do confirm this fact ($f > 1$), then this kind STG might be a useful candidates to explain it.

We show the current observational limits on $f$ and $\gamma$ in Table \ref{table2}. $\gamma$ value is obtained from $f^{\rm{obs}}$ when we assume $\Lambda$ model. The current observations may be consistent with $\Lambda$ model except the Lyman-$\alpha$ forest result \cite{0407377}. However, there exist huge errors in observations and it might be too early to extract any physical properties for specific models.

\begin{center}
    \begin{table}
    \begin{tabular}{ | c | c | c | c | c | c | }
    \hline
      $z_{\ast}$ & $n_{\ast}$  & $\Omo$ & $f^{\rm{obs}}$ & $\gamma^{\rm{obs}}$ & Ref \\ \hline
      $0.15$ & $-0.14$ & $0.3$  & $0.51 \pm 0.11$ & $0.72^{+ 0.26}_{- 0.21}$ & 2dFGRS \cite{0112161,0212375} \\ \hline
      $0.32$ & $-0.28$ & $0.26 $  & $0.654_{-0.132}^{+0.185}$ & $0.52^{+0.28}_{-0.31}$ & SDSS R$= 10-50 h^{-1}$ Mpc \cite{10032185} \\ \cline{4-6}
      $ $ &  & $\pm 0.02$ & $0.641_{-0.134}^{+0.191}$ & $0.55^{+0.29}_{-0.32}$ & R$= 2-50 h^{-1}$ Mpc  \\ \hline
      $0.35$ & $-0.3$ & $0.3$  & $0.70 \pm 0.18$ & $0.54^{+ 0.45}_{- 0.34}$ & SDSS \cite{0608632} \\ \hline
      $0.55$ & $-0.44$ & $0.3$ & $0.75 \pm 0.18$ & $0.59^{+ 0.60}_{- 0.44}$ & 2dF-SDSS \cite{0612400} \\ \hline
      $1.4$ & $-0.88$ & $0.3$  & $0.90 \pm 0.24$ & $0.68^{+ 2.0}_{- 1.5}$ &  2dF-SDSS \cite{0612401} \\ \hline
      $3.0$ & $-1.39$ & $0.3$  & $1.46 \pm 0.29$ & $-10.6^{+ 6.2}_{- 5.1}$ & Ly-$\alpha$ (SDSS) \cite{0407377} \\ \hline
    \end{tabular}
    \caption{$z_{\ast}$ is the corresponding redshift for each observation. $\gamma^{\rm{obs}}$ is the derived quantities from the observational value of $f^{\rm{obs}}$ when we assume $\Lambda$ model ($\omega = -1$). }
    \label{table2}
    \end{table}
\end{center}

\section{Conclusions}
\setcounter{equation}{0}

Scalar-tensor gravities theories can produce many possible background evolutions which mimic dark energy models and other modified gravity models. We need to consider the growth of the linear matter perturbation to distinguish between models. However, STG models are strongly limited by the solar system constraint when we normalize the kinetic energy term $Z(\phi) = 1$. The main reason for this is that $\phi'$ becomes singular with this normalization. Thus, we may need to investigate the STG models with general $Z(\phi)$ in order to distinguish STG model with others.

When we allow the time variation of the growth index parameter $\gamma$, usually the negative value of $\gamma_{a}$ models have the singular problem of $\phi'$. Some models with the positive $\gamma_{a}$ have the interesting features like large enough $F/F_0$ values at early epoch while mimic the dark energy models background evolution. However, these cases violate the solar system test and again this might be able to be cured when we consider STG models. If $\gamma_a$ is positive (negative), then $F(n)/F_0$ shows the convex (concave) shape with the minimum (maximum) as $1$. Thus, the value of $F(n)$ is bigger than $F_0$ in the past for the viable STG. There can be exception for this case, when $\gamma_a \simeq 0$ but positive.

The main conclusion is that the viable STG models with $Z(\phi) = 1$ are not distinguishable from dark energy models or other modified gravity models when we strongly limit the solar system constraint.

\appendix
\section{Appendix}
\setcounter{equation}{0}

\begin{center}
\begin{figure}
\vspace{1.5cm}
\centerline{\psfig{file=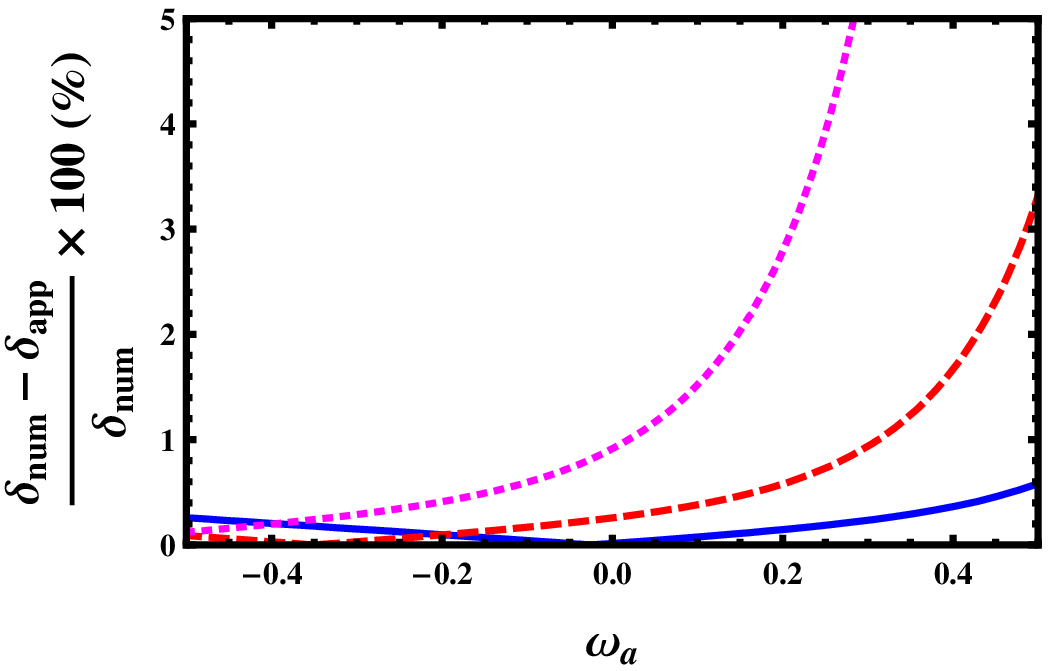, width=6.5cm} \psfig{file=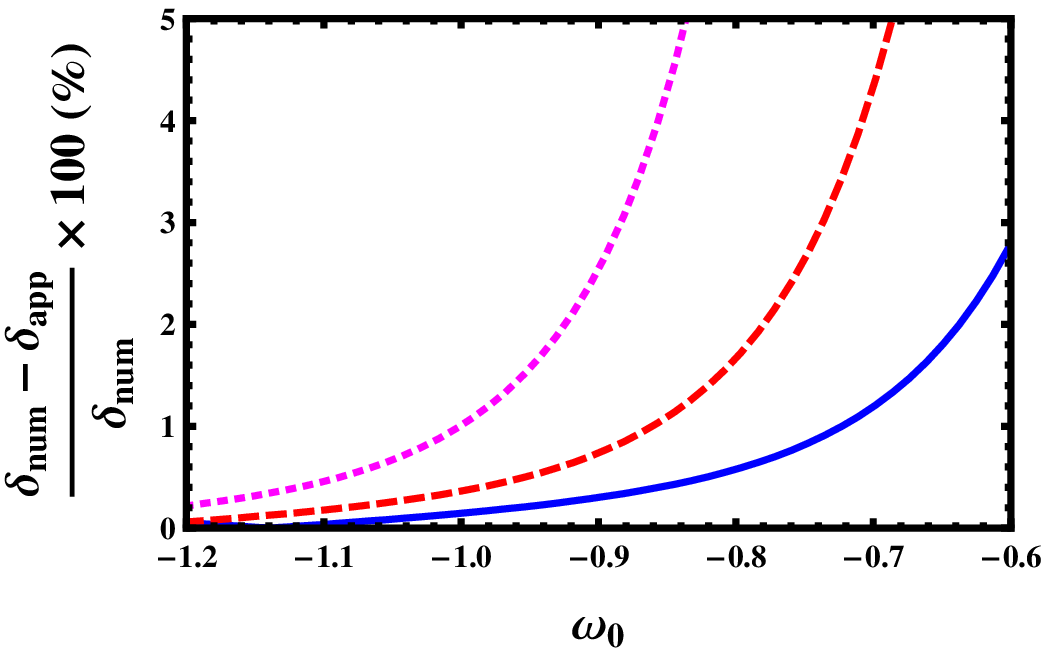, width=6.7cm}}
\vspace{-0.5cm}
\caption{ a) Errors in the growth factor $\delta$ obtained from Eq. (\ref{Linderg}) compared to the exact (numerical) solution as a function of $\oa$ for the different values of $\oo = -1.0$, $-0.8$, and $-0.6$ (from bottom to top). b) Errors as a function of $\oo$ for the different values of $\oa = 0.2$, $0.4$, and $0.6$ (from left to right).} \label{fig4}
\end{figure}
\end{center}

Although the functional form of the growth factor given in Ref. \cite{0507263} is simple and useful, its usage should be limited for certain values of $\omega$. If $\oo$ and (or) $\oa$ are (is) big, the form loses the accuracy. We show this in Fig. \ref{fig4}. We show the errors in $\delta$ for the different values of $\oo = -1.0$ (solid), $-0.8$ (dashed), and $-0.6$ (dotted) as a function of $\oa$ in the left panel of Fig. \ref{fig4}. Except for $\oo = -1.0$, the errors increase rapidly as $\oa$ increases. We also probe the errors for the different values of $\oa = 0.2$ (solid), $0.4$ (dashed), and $0.6$ (dotted) as a function of $\oo$ in the right panel of Fig. \ref{fig4}. The errors are more than $1$ \% in many cases. Thus, one should not rely on the functional form for general case.

Also this functional form does not depend on $\Omo$. Thus, if one study some model with small or large enough values of $\Omo$, then one should not rely on this functional form. Instead one can improve $\gamma$ by using the exact solution of $\delta$ as shown in Refs. \cite{09051522,09061643,09072108} \be \gamma = \ln \Biggl[ 1 - \fr{3 Q ( -1 + \omeff) F \Bigl[1 - \fr{1}{3 \omeff}, \fr{3}{2} - \fr{1}{2 \omeff}, 2 - \fr{5}{6 \omeff}, - Q \Bigr]}{(-5 + 6 \omeff) F \Bigl[ - \fr{1}{3 \omeff}, \fr{1}{2} - \fr{1}{2 \omeff}, 1 - \fr{5}{6 \omeff}, -Q \Bigr]} \Biggr] \Biggl/ \ln [\Omo] \label{gammaL} \ee where $Q = \fr{1 - \Omo}{\Omo}$ and $\omeff = \omega_{0} + \fr{\omega_{a}}{2}$.

The initial values of $\phi^{'}$ and $U(\phi)$ are determined if we know the exact values of ($\oo, \oa$), ($\gamma_o, \gamma_a$), and $\Omega_{\m}^{0}$, which might be possible up to certain level in future observations. We use Eqs. (\ref{G00n}), (\ref{omde}), (\ref{FoF0}), (\ref{FpoF0}), and (\ref{FppoF0}) to obtain $\phi^{'}_{0}$ and $U(\phi_0)$. \ba \fr{\phi_{0}^{'2}}{2 F_0} &=& - \fr{F_0^{''}}{3 F_0} + \Bigl(4 - \fr{H_{0}'}{H_0} \Bigr) \fr{F_0'}{3 F_0} + ( 2 + \oo ) ( 1 - \Omo) \, , \label{phi0p} \\ \fr{U_0}{F_0 H_0^2} &=& \fr{F_0''}{3F_0} + \Bigl( 5 + \fr{H_0'}{H_0} \Bigr) \fr{F_0'}{3 F_0} + ( 1 - \oo ) (1 - \Omo) \, , \label{U0} \ea where \ba \fr{F_0'}{F_0} &=& - \Biggl( 3 + 2 \fr{H_0'}{H_0} + \fr{P_0'}{P_0} \Biggr) \, , \label{Fp0oF0} \\ \fr{F_0''}{F_0} &=& \Biggl( 3 + 2 \fr{H_0'}{H_0} + \fr{P_0'}{P_0} \Biggr)^2 - 2 \Biggl( \fr{H_0'}{H_0} \Biggr)' - \Biggl( \fr{P_0'}{P_0} \Biggr)' \, , \label{Fpp0oF0} \\ P_0 &=& \Bigl( \Omo^{\gamma0} \Bigr)' + \Omo^{2\gamma0} + \Omo^{\gamma0} \Biggl( 2 + \fr{H_0'}{H_0} \Biggr) \, , \label{P0} \\ P_0' &=& \Bigl( \Omo^{\gamma0} \Bigr)'' + \Bigl( \Omo^{\gamma0} \Bigr)' \Biggl( 2 \Omo^{\gamma0} + 2 + \fr{H_0'}{H_0} \Biggr) + \Omo^{\gamma0} \Biggl( \fr{H_0'}{H_0} \Biggr)' \, , \label{P0p} \\ P_0'' &=& \Bigl( \Omo^{\gamma0} \Bigr)''' + \Bigl( \Omo^{\gamma0} \Bigr)'' \Biggl( 2 \Omo^{\gamma0} + 2 + \fr{H_0'}{H_0} \Biggr) + 2 \Bigl( \Omo^{\gamma0} \Bigr)' \Biggl( \Omo^{\gamma0} + \fr{H_0'}{H_0} \Biggr)' + \Omo^{\gamma0} \Biggl(\fr{H_0'}{H_0} \Biggr)'' \, , \label{P0pp} \\ \fr{H_{0}'}{H_0} &=& -\fr{3}{2} \Bigl[ 1 + \oo (1 - \Omo) \Bigr] \equiv - \fr{3 + Q_0}{2} \, , \label{Hp0oH0} \\ \Biggl( \fr{H_0'}{H_0} \Biggr)' &=& \fr{3}{2} \Bigl[ \oa ( 1 - \Omo) + 3 \oo^2 \Omo ( 1 - \Omo) \Bigr] \equiv -\fr{Q_0'}{2} \, , \label{Hp0oHoP} \\ \Biggl( \fr{H_0'}{H_0} \Biggr)'' &=& \fr{3}{2} \Bigl[ \oa ( 1 - \Omo) - 9 \oo \oa \Omo ( 1 - \Omo) - 9 \oo^3 \Omo ( 1 - \Omo) (2 \Omo - 1) \Bigr] \nonumber \\ &\equiv& -\fr{Q_0''}{2} \, , \label{Hp0oHoPP} \\ \Bigl( \Omo^{\gamma0} \Bigr)' &=& \Omo^{\gamma0} \Bigl( \gamma_0 Q_0 - \gamma_a \ln \Omo \Bigr) \, , \label{Omogam0P} \\ \Bigl( \Omo^{\gamma0} \Bigr)'' &=& \Omo^{\gamma0} \Biggl( \Bigl[ \gamma_0 Q_0 - \gamma_a \ln \Omo \Bigr]^2 + \Bigl[ -\gamma_a \ln \Omo - 2 \gamma_a Q_0 + \gamma_0 Q_0' \Bigr] \Biggr) \, , \label{Omogam0PP} \\ \Bigl( \Omo^{\gamma0} \Bigr)''' &=& \Omo^{\gamma0} \Biggl( \Bigl[ \gamma_0 Q_0 - \gamma_a \ln \Omo \Bigr]^3 + 3 \Bigl[ \gamma_0 Q_0 - \gamma_a \ln \Omo \Bigr] \Bigl[ -\gamma_a \ln \Omo - 2 \gamma_a Q_0 + \gamma_0 Q_0' \Bigr] \nonumber \\ && +  \Bigl[ -\gamma_a \ln \Omo - 3 \gamma_a Q_0 - 3 \gamma_a Q_0' + \gamma_0 Q_0'' \Bigr] \Biggr) \, , \label{Omogam0PPP} \ea

\section*{Acknowledgments}
This work was supported in part by the National Science Council, Taiwan, ROC under the
Grant NSC 98-2112-M-001-009-MY3.

\end{document}